\documentclass[a4paper,twoside,10pt]{letter}
\usepackage{graphicx,saj,multicol ,subeqnarray}

\newcommand{\D}{$^\circ$}

\newcommand{\HII}{H\,{\sc ii}}

\newcommand{\SII}{[S\,{\sc ii}]}

\newcommand{\OII}{[O\,{\sc ii}]}
\newcommand{\OIII}{[O\,{\sc iii}]}

\newcommand{\HA}{H{$\alpha$}}

\newcommand{\HB}{H{$\beta$}}

\newcommand{\msun}{M$_{\odot}$}

\def\arcmin{\hbox{$^\prime$}}
\def\arcsec{\hbox{$^{\prime\prime}$}}

\def\papertype{}

\def\udc{...}
\begin{document}
\baselineskip=3.1truemm
\columnsep=.5truecm
\newenvironment{lefteqnarray}{\arraycolsep=0pt\begin{eqnarray}}
{\end{eqnarray}\protect\aftergroup\ignorespaces}
\newenvironment{lefteqnarray*}{\arraycolsep=0pt\begin{eqnarray*}}
{\end{eqnarray*}\protect\aftergroup\ignorespaces}
\newenvironment{leftsubeqnarray}{\arraycolsep=0pt\begin{subeqnarray}}
{\end{subeqnarray}\protect\aftergroup\ignorespaces}
%

% Running titles

\markboth{\eightrm SMC~SMP~24: A NEWLY RADIO-DETECTED PLANETARY NEBULA IN THE SMALL MAGELLANIC CLOUD}
{\eightrm  I.~S.~BOJI\v{C}I\'C, M.~D.~FILIPOVI\'C and  E. ~J.~Crawford}

{\ }

\publ

\type

{\ }

% Title

\title{SMC~SMP~24: A NEWLY RADIO-DETECTED PLANETARY NEBULA IN THE SMALL MAGELLANIC CLOUD}

% Authors

\authors{I.~S.~Boji\v{c}i\'c, M.~D.~Filipovi\'c and  E. ~J.~Crawford}

\vskip3mm

% Address

\address{University of Western Sydney, Locked Bag 1797, Penrith South DC, NSW 1797, Australia}
\Email{i.bojicic}{uws.edu.au}

% Received and Accepted dates

\dates{September XX, 2010}{September XX, 2010}

% Abstract

\summary{In this paper we report new radio-continuum detection of an extragalactic PN: SMC~SMP~24. We show the radio-continuum image of this PN and present the measured radio data. The newly reduced radio observations are consistent with the multi-wavelength data and derived parameters found in the literature. SMC~SMP~24 appear to be a young and compact PN, optically thick at frequencies below 2~GHz.}

% Keywords (see keywords.pdf file)

\keywords{ISM: planetary nebulae -- Magellanic Clouds -- Radio
  Continuum: -- ISM: individual objects -- SMC~SMP~24}

\begin{multicols}{2}
{

% Sections

\section{1. INTRODUCTION}

The importance of the radio-continuum properties of planetary nebulae (PNe) has been recently reinstated with the report of the radio-continuum observations of PNe in the Magellanic Clouds (Filipovi\'c et al., 2009). The comprehensive multifrequency study, based on Australia Telescope Compact Array+Parkes mosaic surveys of Magellanic Clouds (MCs) (Hughes et al. 2006, 2007; Payne et al. 2004, 2009; Filipovi\'c et al. 1995, 1997, 2002, 2005, 2008), helped to reveal the true nature of more than 50 PN candidates in MCs as compact \HII\ regions. Also, based on our radio data, Reid \& Parker (2010) were able to re-classify three ultra-bright PNe (previously classified as 'true' PNe) as contaminants due to their strong radio fluxes. Our MCs radio PNe detections represent only $\sim$3 per cent of the optical PNe population of the MCs. Most likely, we are selecting only the strongest radio-continuum emitters, possibly at a variety of different stages of their evolution (Vukoti\'c et al. 2009).

Prior to this study, a radio detection of only three extragalactic PNe have been reported in the literature (Zijlstra et al., 1994; Dudziak et al., 2000).  Based on the radio-continuum properties of radio-bright Galactic PNe the expected radio flux densities at the distance of the Large Magellanic Cloud (LMC) and the Small Magellanic Cloud (SMC) are up to $\sim$2.5 and $\sim$2.0~mJy at 1.4~GHz, respectively. While the LMC sample conform with the radio luminosity limit predicted from Galactic PNe, the SMC PNe sample appear to be unusually strong radio emitters.

The known and well refined distances to the LMC/SMC provide a great opportunity for the accurate evaluation of important physical properties for PNe such as ionised massed and electron densities. Also, a statistically significant sample of radio detected extragalactic PNe will allow the construction and examination of the bright end of the radio PN luminosity function (PNLF) and comparison with established theoretical and empirical MCs PNLF's obtained at other wavelengths (Jacoby et al., 1990; Mendez et al., 1993; Stanghellini, 1995; Jacoby \& De Marco, 2002; Ciardullo et al., 2002). 

Therefore, we initiated a deep, 6~cm radio-continuum survey which will attempt to detect and accurately measure the radio-continuum flux densities of 50+ MCs PNe. As a first step in this project we thoroughly examined the archived Australia Telescope Compact Array (ATCA) data (the Australia Telescope Online Archive\footnote{http://atoa.atnf.csiro.au/}) in order to find if any of the objects have been already observed and to estimate upper flux limits for the prepared sample. In this paper we report the radio detection of SMC~SMP~24 (hereafter SMP~24), found in the ATCA's archival deep observation of the SMC supernova remnant SNR\,0101-7226 conducted in 1993/1994 by Ye et al. (1995). 

\section{2. Multiwavelength data}

SMP~24 (LHA\,115 -- N\,S70; Henize, 1956) is a SMC PN located approximately 2~arcmin north of the N\,S66 giant star-forming complex. From its appearance in \HA\ and \OIII\ spectral lines, Stanghellini et al. (2003) designated this PN as an elliptical with a possible faint halo. The authors reported a flux calibrated intensity in the \HB\ line, optical extinction (c$_{H\beta}$), and relative intensities of several bright spectral lines. Also, the photometric radii of the nebula and the nebular dimensions, measured from the 10\% brightness contour have been reported. They determined electron density of $\log n_{e}\approx3.1$ from \SII$\lambda\lambda$6716,6731 lines and calculated ionised mass of the nebula (M$_{i}\approx0.86$~\msun). It is important to note that the reported ionised mass, calculated using the eq.~6 from Boffi \& Stanghellini (1994), is, by a factor of 2.8 and 4, larger than the average mass of the rest of PNe from the SMC and LMC sample reported in this paper, respectively. From the \OII\ electron density diagnostic (I($\lambda$3726.0)/I($\lambda$3728.8), Stanghellini et al. (2009) and Shaw et al. (2010) reported a significantly higher electron density for this PN of $\log n_{e}\approx3.4$. Assuming that the rest of the parameters used by Stanghellini et al. (2003) stay the same, the new electron density estimate will half the ionised mass to M$_{i}\approx0.4$. The newly obtained mass is in a much better agreement with the rest of the SMC PNe sample. We tabulated the reported data, relevant to this study, in Table~1. 

SMP~24 is observed as a part of spectroscopic observations of 25 MCs PNe conducted with the {\it Spitzer Space Telescope} Infrared Spectrograph (Bernard-Salas et al. 2009). The presence of hydrogenated amorphous carbon molecules (HACs) in this PN, is interpreted by the authors as an evidence of the early evolutionary stage. The SMP~24 central star is characterised by a very low ionisation potential, also noticed by Stanghellini et al. (2009) (EC 1-2). Also, this PN is detected in the Spitzer Survey of the Small Magellanic Cloud (S$^{3}$MC: Bolatto et al. 2007), which imaged the star-forming body of the SMC in all seven MIPS and IRAC wave bands. Measured flux densities from the $B$~band to the 70~$\mu$m band (where the PN is not detected) are tabulated in Table~2. 

%Finally, Oskinova \& Brown (2003)  reported detection of SMP~24 in X-ray band by Chandra ACIS-S.  However, the detection .....

\section{3. 20~cm detection~of~SMP~24}

SMP~24 was observed by T. Ye, S. Amy, L. Ball and J. Dickel with the ATCA as a part of project C281 over two 12 hour sessions on 25$^\mathrm{th}$ August 1993 and 10$^\mathrm{th}$ February 1994. Two complementary array configurations at 20/13~cm ($\nu$=1377/2377~MHz) were used -- 1.5B and 6B. However, SMP~24 is positioned some $\sim$18\arcmin\ from the pointing centre and therefore appeared outside of the primary beam of the 13~cm observations. The source 1934-638 was used for primary calibration and source 0252-712 as the secondary calibrator. More information about observing procedure and other sources observed during these sessions can be found in Ye et al. (1995).

The \textsc{miriad} (Sault and Killeen 2010) and \textsc{karma} (Gooch~2006) software packages were used for reduction and analysis.  The initial, high-resolution image was produced from the full data-set and using \textsc{miriad} multi-frequency synthesis (Sault and Wieringa~1994) and natural weighting. The obtained 20~cm image has a resolution of 7.0\arcsec$\times$6.6\arcsec\ at PA=42.6\D\ and an estimated r.m.s. noise of 0.05 mJy~beam$^{-1}$ which is significantly better than Ye et al. (1995) of 0.1~mJy~beam$^{-1}$. We attribute this difference to slightly different cleaning technique and careful flagging of very noisy observational data. The new high-resolution and high-sensitivity analysis of these observations will be presented in the future papers.

However, due to the effect of the decreasing phase stability with increasing baseline length, which could affect the position and the flux density estimate of faint, point like sources, we created additional ``low-resolution'' total intensity image by excluding long baselines (i.e. without correlations with antenna 6). The excerpts from this ``low-resolution'' map, which is used in this study, are presented in Fig.~1. The map has a resolution of 15.3\arcsec$\times$14.4\arcsec\ at PA=39.8\D\ and an estimated local r.m.s. noise of 0.1~mJy~beam$^{-1}$ measured from the $\sim$8\arcmin$\times$8\arcmin\ box centred on the SMP~24. 

In order to confirm the positional correlation between the newly found radio source and SMP~24 we created a colour composite (RGB) image of the SMP~24 region using data from the Magellanic Cloud Emission Line Survey (MCELS: Smith \& The MCELS Team, 1999). Figure~2 (left) represents the  region of the SMC centred on SMP~24. The PN can be seen in the centre of the field as a distinctive blue point source. In Figure~2 (right) we show a radio-continuum contour map of SMP~24 superposed on the MCELS colour composite image. From Figure~2 can be seen that the peak flux in the radio appear to be well correlated with the peak flux in the optical line emission. 

The position and the peak flux density of the SMP~24 was determined by fitting a two-dimensional Gaussian to the restored and beam-corrected total intensity map. We used \textsc{miriad}'s task {\sc imfit} with a clip level of 5~$\sigma$ (where $\sigma$ is the measured local r.m.s. noise). All pixels below the clipping level were excluded from the fitting process. The error in the measured peak flux density is estimated as a quadrature sum from the locale noise level (0.1~mJy~beam$^{-1}$) and the uncertainty in the gain calibration (10\%). However, due to the non-linear systematic errors which can arise from a large distance from the phase centre and a low signal to noise ratio (e.g. CLEAN bias), we applied additional 30\% of uncertainty in the final flux density estimate.  The results and the parameters used in the fitting procedure are presented in Table~2.

\section{4. Discussion}

The presented multi-wavelength data is in good agreement with suggestion that  SMP~24 is a very young PN (Bernard-Salas et al., 2009). The 20~cm peak flux density observed in this PN corresponds to a low limit for the radio surface brightness temperature of $\sim4.3\times10^{3}$~K (assuming upper limit for the angular diameter of $\theta\approx0.38$). This implies optically thick or at least partially optically thick radio-continuum emission  at 20~cm. The same implication can also arise from a comparison between the measured radio flux density and the flux derived from the \HB\ emission line.  Due to the same dependence on the nebular density it is expected that the centimeter radio-continuum emission and Balmer lines emission will be well correlated (see Pottasch, 1984 eq. IV-26). From parameters tabulated in Tab.~1 (F(\HB), c(\HB), T$_e$ and  $n(\textrm{He}^{+})/n(\textrm{H}^{+})$) we estimated flux at 1.4~GHz of 1.1~mJy. Although within uncertainty range, this 40\% deviation from the measured 1.4~GHz radio-continuum flux indicate that the self-absorption by the nebula is important in this frequency band and implies an existence of significant density stratification (where the reported $\log n_{e}\approx3.4$~cm$^{-3}$ is probably only the average). 

The empirical and modelled spectral energy distributions (SED) of SMP~24 are presented in Figure~3. Since only one observation is available in the radio-continuum band we roughly estimated the position of the turnover (critical) frequency using the obtained brightness temperature at 20~cm. From T$_{b}$=T$_{e}$(1-$e^{-\tau_{\nu}}$) we found an average optical depth at 20~cm of $\tau_{20cm}=0.47$. The critical frequency ($\nu_{c}$) can now be found from: $\tau_{\nu}=(\nu/\nu_{c})^{-2.1}$  (Pottasch, 1984). From these two point we constructed the SED for SMP~24 using the simple approximation of the uniformly ionised region with constant density and constant electron temperature (see eq.~4 in Sharova, 2002). 

Assuming that SMP~24 is a young planetary nebula, we expect that its central star will be still closely surrounded by the ejected dusty envelope. In order to estimate the dust temperature (T$_{d}$) we fitted a blackbody spectrum to the far IR (FIR) data. It is important to note that the R$_{phot}$ measurement from Stanghellini et al. (2003) is used as an estimate for angular diameter of the emitting dust. Also, no attempt was made to estimate the optical depth in IR bands. The best fit is obtained with  T$_{d}\approx270$~K (Fig.~3: dashed line). However, in order to better reproduce the observed SED down to 1~$\mu$m we fitted additional hot dust component in the mid and near IR bands (MIR and NIR, respectively) with estimated T$_{hd}\approx1000$~K (dot-dashed line) and with the same approximations used for the FIR band. The summed SED of the radio-continuum, dust and the hot dust was plotted in Fig~3 with solid line.  The MIR to radio ratio for this object is found to be $\sim$12 which is in accord with a range of values expected for PNe (Cohen \& Green, 2001; Cohen et al., 2007). 

For comparison we overplotted the observed radio-continuum to IR SED of young Galactic PN IC~418 scaled to the distance of SMC. Observational data is from Meixner et al. (1996); Guzm{\'a}n et al. (2009); Pazderska et al. (2009) and Vollmer et al. (2010) and with adopted distance to this PN of 1.3~kpc (Guzm{\'a}n et al., 2009). This object is already used by  Bernard-Salas et al., (2009) for a comparison with SMC PNe SiC feature (broad feature from 9 to 13 $\mu$m seen regularly in carbon stars but very rarely in Galactic PNe). IC~418 is a bright and young C-rich Galactic PN with well defined ring structure with angular diameter of $\sim$12~arcsec ($\sim$0.25~arcsec scaled to the SMC distance). It is surrounded with a low-level ionised halo and embedded in a large molecular envelope (Taylor et al., 1989; Hyung et al., 1994).  As can be seen from Fig.~3 IC~418 shows a clear similarity in SED with SMP~24. 

\section{5. Summary}

In this paper we presented detection of the radio-continuum emission from the SMC PN: SMP~24. This object is a radio luminous PN with estimated flux density at 1.4~GHz of 0.7$\pm$0.4~mJy ($\sim$2.7~Jy scaled to the distance of 1~kpc). Because of the relatively high brightness temperature at 1.4~GHz and the significant difference between the measured radio-continuum flux and predicted from \HB\ we conclude that the ionised shell of this PN is very likely optically thick (or partially optically thick) at frequencies below 2~GHz. However, it is important to note that in order to properly examine radio-continuum properties of this PN, the additional, multi-frequency radio-continuum data is needed. This PN is scheduled to be observed in our ATCA-based follow-up observations of MC's PNe.

We discussed the evolutionary stage and spectral energy distribution of this SMC PN in the light of the available multi-wavelength data and from the evident similarities with a young and well studied Galactic PN IC~418. SMP~24 appear to be a young PN with a dynamic age of $<$1000~yr. The ionised gas and the hot dust are very likely still located in the same region, close to the central star. We believe that our future, high resolution and high sensitivity radio-continuum observations of SMP~24 will help to reveal some of its intrinsic physical properties (e.g. emission measure, physical size and mass of the ionised shell).

% Acknowledgements

\acknowledgements{We used the {\sc karma} and \textsc{miriad} software package developed by the ATNF. The Australia Telescope Compact Array (/ Parkes telescope / Mopra telescope / Long Baseline Array) is part of the Australia Telescope which is funded by the Commonwealth of Australia for operation as a National Facility managed by CSIRO. We thank the Magellanic Clouds Emission Line Survey (MCELS) team for access to the optical images. This research has made use of the SIMBAD database and VizieR catalogue access tool, operated at CDS, Strasbourg, France. We were granted observation time at the South African Astronomical Observatory (SAAO) and wish to thank them for their kind help and accommodations. Travel to the SAAO was funded by Australian Government ANSTO AMNRF grant number 09/10-O-03. This research has been supported by the University of Western Sydney research grant (project number 20721.80758). 
%We thank the referee for numerous helpful comments that have greatly improved the quality of this paper.
}

% References

\references

Bernard-Salas, J., Peeters, E., Sloan, G. C., Gutenkunst, S., Matsuura, M., Tielens, A. G. G. M., Zijlstra, A. A. and Houck, J. R.: 2009, \journal{Astrophys. J.}, \vol{699}, 1541.

Boffi, F. R. and Stanghellini, L.: 1994, \journal{Astron. Astrophys.}, \vol{284}, 248.

Bolatto, A. D., Simon, J. D., Stanimirovi\'c, S., van Loon, J. T., Shah, R. Y., Venn, K., Leroy, A. K., Sandstrom, K., Jackson, J. M., Israel, F. P., Li, A., Staveley-Smith, L., Bot, C., Boulanger, F. and Rubio, M.: 2007, \journal{Astrophys. J.}, \vol{655}, 212.

Ciardullo, R., Feldmeier, J. J., Jacoby, G. H., Kuzio de Naray, R., Laychak, M. B. and Durrell, P. R.: 2002, \journal{Astrophys. J.}, \vol{577}, 311.

Cohen, M. and Green, A. J.: 2001, \journal{Mon. Not. R. Astron. Soc.}, \vol{325}, 531.

Cohen, M., Parker, Q. A., Green, A. J., Murphy, T., Miszalski, B., Frew, D. J., Meade, M. R., Babler, B., Indebetouw, R., Whitney, B. A., Watson, C., Churchwell, E. B. and Watson, D. F.: 2007, \journal{Astrophys. J.}, \vol{669}, 343.

Dudziak, G., P\'equignot, D., Zijlstra, A. A. and Walsh, J. R.: 2000, \journal{Astron. Astrophys.}, \vol{363}, 717.

Filipovi\'c, M. D., Haynes, R. F., White, G. L., Jones, P. A., Klein, U., Wielebinski R.: 1995, \journal{Astron. Astrophys. Journ. Suppl. Series}, \vol{111}, 311.

Filipovi\'c, M. D., Jones, P. A., White, G. L.,Haynes, R. F., Klein, U., Wielebinski, R.: 1997, \journal{Astron. Astrophys. Journ. Suppl. Series}, \vol{121}, 321.

Filipovi\'c, M. D., Bohlsen, T., Reid, W., Staveley-Smith, L., Jones, P. A., Nohejl, K., Goldstein, G.: 2002, \journal{Mon. Not. R. Astron. Soc.}, \vol{335}, 1085.

Filipovi\'c, M. D., Payne, J. L., Reid, W., Danforth, C. W., Staveley-Smith, L., Jones, P. A., White, G. L.: 2005, \journal{Mon. Not. R. Astron. Soc.}, \vol{364}, 217.

Filipovi\'c M. D., et al.: 2008, in van Loon J. T., Oliveira J.M., eds, IAU Symp. \vol{256}, The Magellanic System: Stars, Gas, and Galaxies. Cambridge Univ. Press, Cambridge, p. PDFÐ8.

Filipovi\'c, M. D., Cohen, M., Reid, W. A., Payne, J. L., Parker, Q. A., Crawford, E. J., Boji\v{c}i\'c, I. S., de Horta, A. Y., Hughes, A., Dickel, J. and Stootman, F.: 2009, \journal{Mon. Not. R. Astron. Soc.}, \vol{399}, 1098.

Gooch, R., 1996, in G. H. Jacoby \& J. Barnes ed., Astronomical Data Analysis Software and Systems, \vol{101} of \journal{Astronomical Society of the Pacific Conference Series}, p. 80.

Guzm\'an, L., Loinard, L., Go\'mez, Y. and Morisset, C.: 2009, \journal{Astron. J.}, \vol{138}, 46.

Henize, K. G.: 1956, \journal{Astrophys. J., Suppl. Ser.}, \vol{2}, 315.

Hyung, S., Aller, L. H. and Feibelman, W. A.: 1994, \journal{Publ. Astron. Soc. Pac.}, \vol{106}, 745.

Hughes, A., Staveley-Smith, L., Kim, S., Wolleben, M., Filipovi\'c M.: 2007, \journal{Mon. Not. R. Astron. Soc.}, \vol{382}, 543.

Hughes, A., Wong, T., Ekers, R., Staveley-Smith, L., Filipovi\'c, M., Maddison, S., Fukui, Y., Mizuno, N.: 2006, \journal{Mon. Not. R. Astron. Soc.}, \vol{370}, 363.

Idiart, T. P., Maciel, W. J. and Costa, R. D. D.: 2007, \journal{Astron. Astrophys.}, \vol{472}, 101.

Jacoby, G. H., Ciardullo, R. and Walker, A. R.: 1990, \journal{Astrophys. J.}, \vol{365}, 471.

Jacoby, G. H. and De Marco, O.: 2002, \journal{Astron. J.}, \vol{123}, 269.

Meixner, M., Skinner, C. J., Keto, E., Zijlstra, A., Hoare, M. G., Arens, J. F. and Jernigan, J. G.: 1996, \journal{Astron. Astrophys.}, \vol{313}, 234.

Mendez, R. H., Kudritzki, R. P., Ciardullo, R. and Jacoby, G. H.: 1993, \journal{Astron. Astrophys.}, \vol{275}, 534.

Oskinova, L. and Brown, J. C.: 2003, in S. Kwok, M. Dopita, \& R. Sutherland ed., Planetary Nebulae: Their Evolution and Role in the Universe \vol{209} of IAU Symposium, p. 425.

Payne, J. L., Filipovi\'c, M. D., Reid, W., Jones, P. A., Staveley-Smith, L., White,
G. L.: 2004, \journal{Mon. Not. R. Astron. Soc.}, \vol{355}, 44.

Payne, J. L., Tauber, L. A., Filipovi\'c, M. D., Crawford, E. J., de Horta, A.: 2009, \journal{Serb. Astron. J.}, \vol{178}, 65

Pazderska, B. M., Gawron\'ski, M. P., Feiler, R., Birkinshaw, M., Browne, I. W. A., Davis, R., Kus, A. J., Lancaster, K., Lowe, S. R., Pazderski, E., Peel, M. and Wilkinson, P. N.: 2009, \journal{Astron. Astrophys.}, \vol{498}, 463.

Pottasch, S. R.: 1984, in: Planetary nebulae - A study of late stages of stellar evolution. \vol.{107} of Astrophysics and Space Science Library, D. Reidel.

Reid, W. A. and Parker, Q. A.: 2010, \journal{Mon. Not. R. Astron. Soc.}, \vol{405}, 1349.

Sharova, O. I.: 2002, \journal{Astronomical and Astrophysical Transactions}, \vol{21}, 271.

Shaw, R. A., Lee, T., Stanghellini, L., Davies, J. E., Garc\'ia-Hern\'andez, D. A., Garc\'ia-Lario, P., Perea-Calder\'on, J. V., Villaver, E., Manchado, A., Palen, S. and Balick B.: 2010, \journal{Astrophys. J.}, \vol{717}, 5622.

Smith, R. C., Team, T. M.: 1999, in Y.-H. Chu, N. Suntzeff, J. Hesser, \& D. Bohlender ed., New Views of the Magellanic Clouds \vol{190} of IAU Symposium, p. 28.

Sault, R.~J., Killeen, N.: 2010, Miriad Users Guide, ATNF, Sydney.

Sault, R.~J., Wieringa, M.~H.: 1994, \journal{Astron. Astrophys. Suppl. Series}, \vol{108}, 585.

Stanghellini, L.: 1995, \journal{Astrophys. J.}, \vol{452}, 515.

Stanghellini, L., Lee, T.-H., Shaw, R. A., Balick, B. and Villaver, E.: 2009, \journal{Astrophys. J.}, \vol{702}, 733.

Stanghellini, L., Shaw, R. A., Balick, B., Mutchler, M., Blades, J. C. and Villaver, E.: 2003, \journal{Astrophys. J.}, \vol{596}, 997.

Taylor, A. R., Gussie, G. T. and Goss, W. M.: 1989, \journal{Astrophys. J.}, \vol{340}, 932.

Vollmer, B., Gassmann, B., Derri\`ere, S., Boch, T., Louys, M., Bonnarel, F., Dubois, P., Genova, F., Ochsenbein, F.: 2010, \journal{Astron. Astrophys.}, \vol{511}, A53.

%	Vollmer B., 2009, VizieR Online Data Catalog, \vol{8085}, 0

Vukoti\'c, B., Uro\v{s}evi\'c, D., Filipovi\'c, M. D., Payne, J. L.: 2009, \journal{Astron. Astrophys.}, \vol{503}, 855.

Ye, T. S., Amy, S. W., Wang, Q. D., Ball, L. and Dickel, J.: 1995, \journal{Mon. Not. R. Astron. Soc.}, \vol{275}, 1218.

Zijlstra, A. A., van Hoof, P. A. M., Chapman, J. M. and Loup, C.: 1994, \journal{Astron. Astrophys.}, \vol{290}, 228.

\endreferences

}
\end{multicols} 

\newpage

\vskip.5cm
\centerline{{\bf Table 1.} SMP~24: Multi-wavelength data and parameters compiled from the literature.}
\vskip2mm
\centerline{\begin{tabular}{lccc}
\hline
Parameter &  &  & Reference\\
\hline
\smallskip
RA(2000)							&00 59 16.6				&					& 1 \\
\smallskip 
DEC(2000)						&-72 02 00.8				&					& 1 \\	
\smallskip
$\log{\textrm{F(\HB}:\lambda4861)}$ 				& -12.66					&($\frac{ergs}{cm^{2} s}$)	& 2	\\
\smallskip
c(\HB)							& 0.047					&					& 2	\\
\smallskip
R$_{phot}$ 						& 0.20					&(arcsec)				& 2	\\
\smallskip
$\theta$							& 0.38					&(arcsec)				& 2	\\
\smallskip
T$_{e}$ 							& 11620$^{+910}_{-740}$K	&(K)					& 3	\\
\smallskip
$\log n_{e}$ 						& 3.4 (3.1$^{\dagger}$)		&(cm$^{-3}$)			& 3(2)	\\
\smallskip
$M_{ion}$	 						& 0.4 (0.86$^{\dagger}$)		&(\msun)				& 4(2)	\\
\smallskip
$n(\textrm{He}^{+})/n(\textrm{H}^{+})$	& 0.097$\pm$0.011			&					& 5	\\
\hline
\end{tabular}}
{\small 
\noindent References: 1) Jacoby \& De Marco (2002), 2) Stanghellini et al. (2003), 
3) Shaw et al. (2010), 4) this paper, 5) Idiart et al. (2007)}
\vskip.5cm

\centerline{{\bf Table 2.} SMP~24: IR data compiled from the literature.}
\begin{footnotesize}
\vskip2mm
\centerline{\begin{tabular}{ccccccccccccc}
\hline
Band & $B$ & $V$ & $I$ & $J$ & $H$ & $K$ & 3.6~$\mu$m & 4.5~$\mu$m & 5.8~$\mu$m & 8.0~$\mu$m & 24~$\mu$m & 70~$\mu$m \\
\hline
F$_{\lambda}$(mJy) &0.878 & 1.622 & 0.395 & 0.473 & 0.534 & 0.902 & 1.870 & 2.252 & 3.813 & 9.292 & 28.028 & $<$(10$\times$)5\\
\hline
\end{tabular}}
\vskip.5cm
\end{footnotesize}

\vskip.5cm
\centerline{{\bf Table 3.} SMP~24: ATCA radio-continuum data.}
\vskip2mm
\centerline{\begin{tabular}{lrc}
\hline
Parameter &  &  \\
\hline
\smallskip
Frequency						&1337					& MHz				\\
\smallskip 
Synth. Beam						&15$\times$14				& (arcsec)				\\	
\smallskip
local r.m.s. noise ($\sigma$)			& 0.1						& (mJy beam$^{-1}$)		\\
\smallskip
Peak Flux							& 0.73$\pm$0.13			& (mJy beam$^{-1}$)		\\
\smallskip
Flux Density						& 0.7$\pm$0.4				& (mJy)				\\				
\smallskip
RA(2000)	 						& 00 59 16.3				&					\\
\smallskip
DEC(2000)						& -72 01 59.9				&					\\
\hline
\end{tabular}}
%{\small 
%\noindent $^{a}$ Measured in 5$\times$5~arcmin box centred on the SMP~24.\\
%$^{b}$ Measured using the \textsc{miriad}`s task IMFIT.\\}
\vskip.5cm

\newpage

\centerline{\includegraphics[angle=-90,scale=0.34]{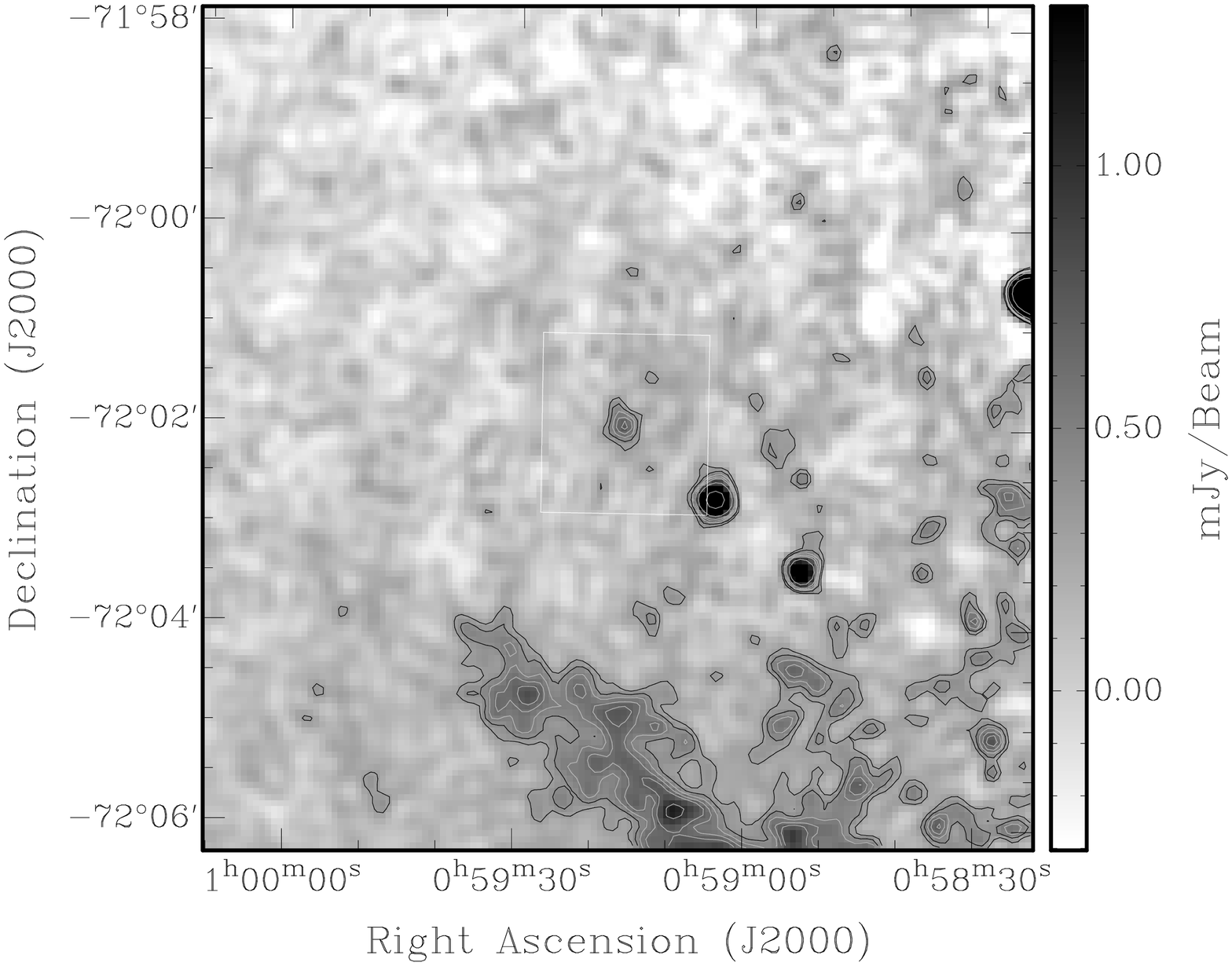}
\includegraphics[angle=-90,scale=0.34]{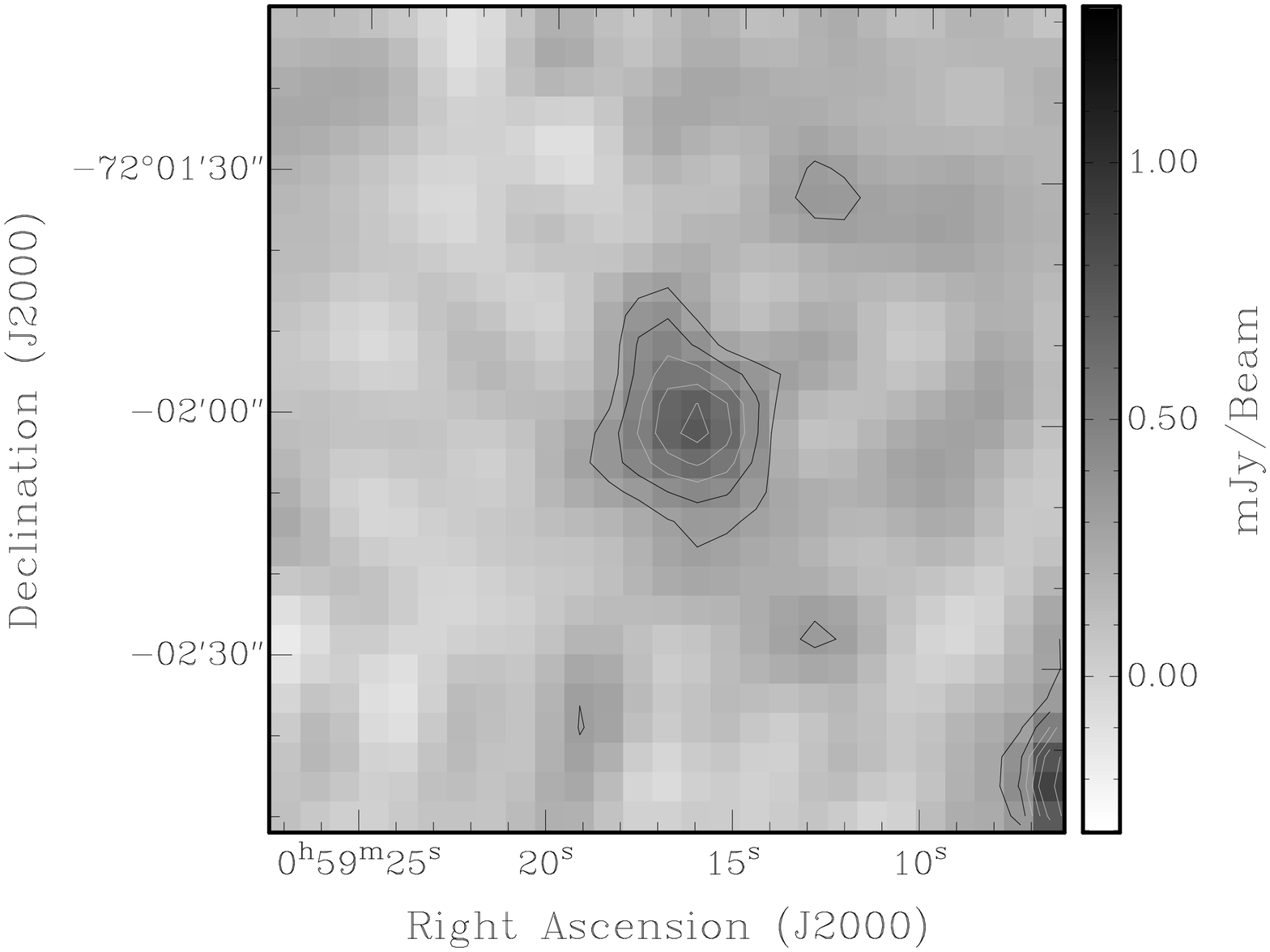}}
\figurecaption{1.}{A radio-continuum total intensity images of SMP~24 overlaid with contours at: 0.3, 0.4 (black), 0.5, 0.6, 0.7, 1 and 2~mJy (grey). The white box in the left panel indicates the zoom region which is presented on the right. The beam size is shown in the bottom left corner of each of the images. }

\centerline{\includegraphics[angle=-90,scale=0.36]{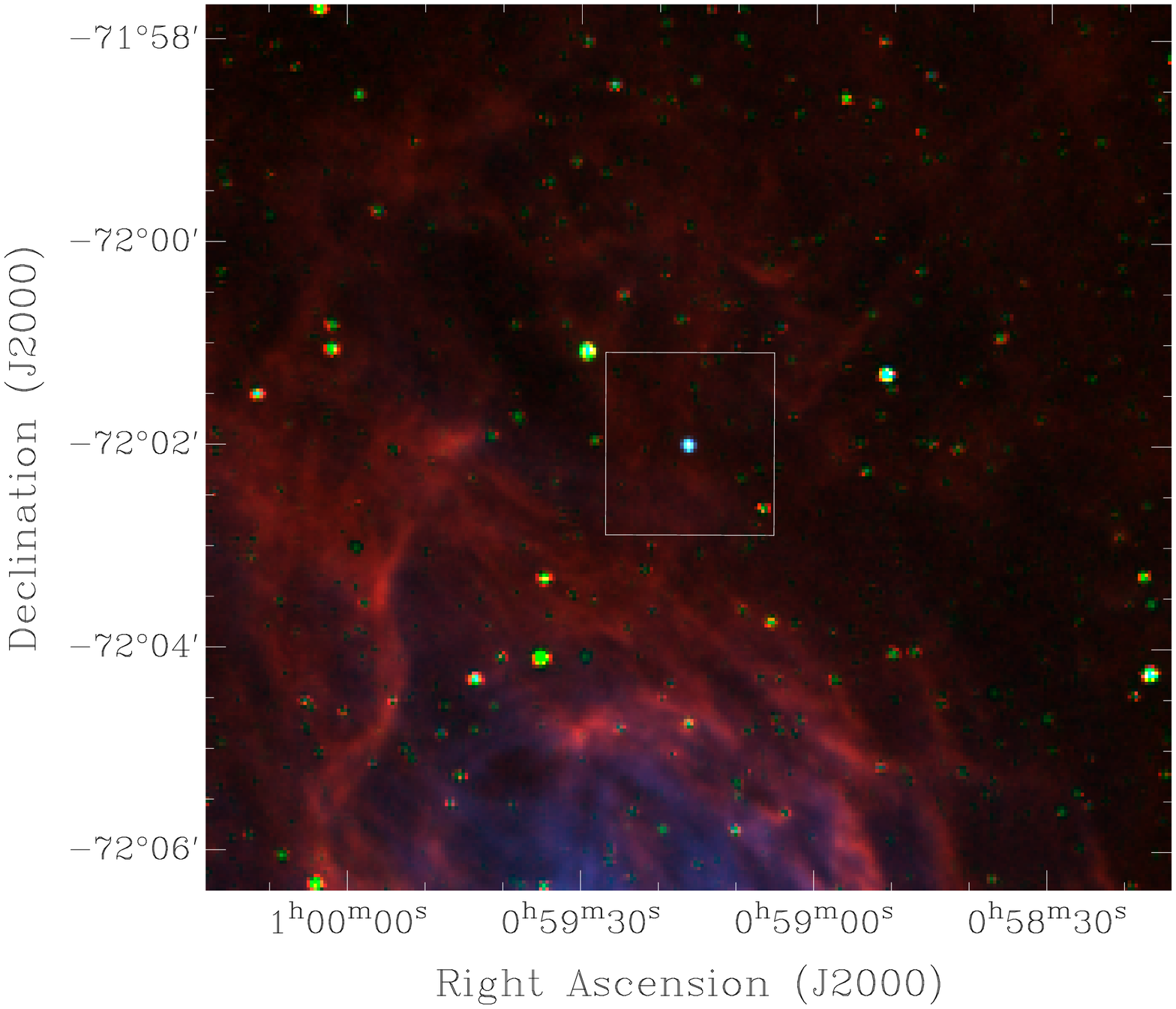}
\includegraphics[angle=-90,scale=0.36]{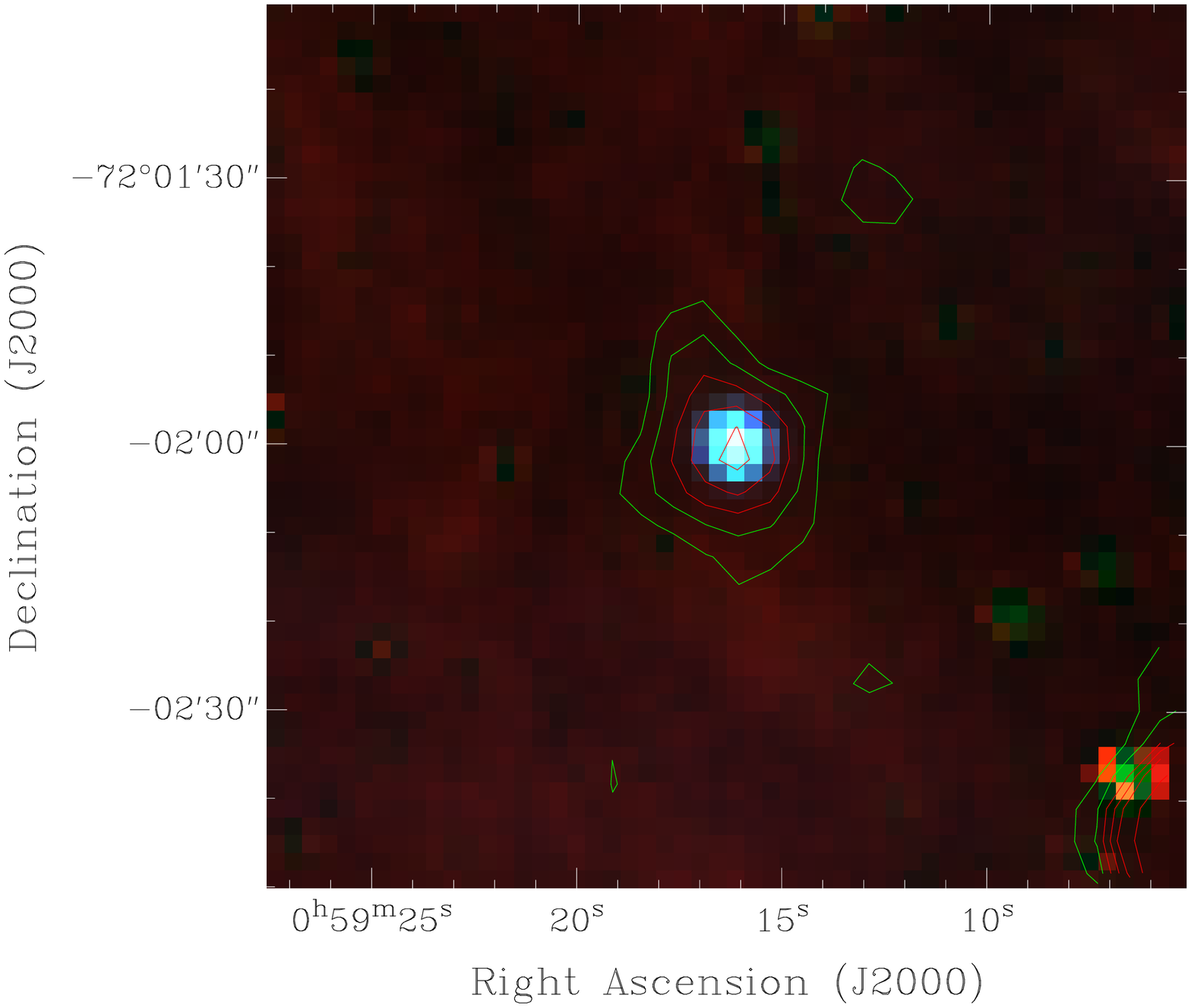}}
\figurecaption{2.}{Colour composite (RGB) images of the SMP~24 region with \HA\ in green, \OIII\ in blue and \SII\ in red, and with arbitrary intensity scaling. {\bf Left}: The 8$\times$8~arcmin region of the SMC where SMP~24 is located. The PN can be seen in the centre of the field as a distinctive blue point source. The white box indicate the ``zoom'' region which is presented on the right. The part of the star forming region N~66 can be seen to the South of the PN. {\bf Right}: A radio-continuum contour map of SMP~24 superposed on the MCELS colour composite image. Contours are at: 0.3, 0.4 (dark green), 0.5, 0.6, 0.7, 1 and 2~mJy (red). }

\centerline{\includegraphics[height=6cm]{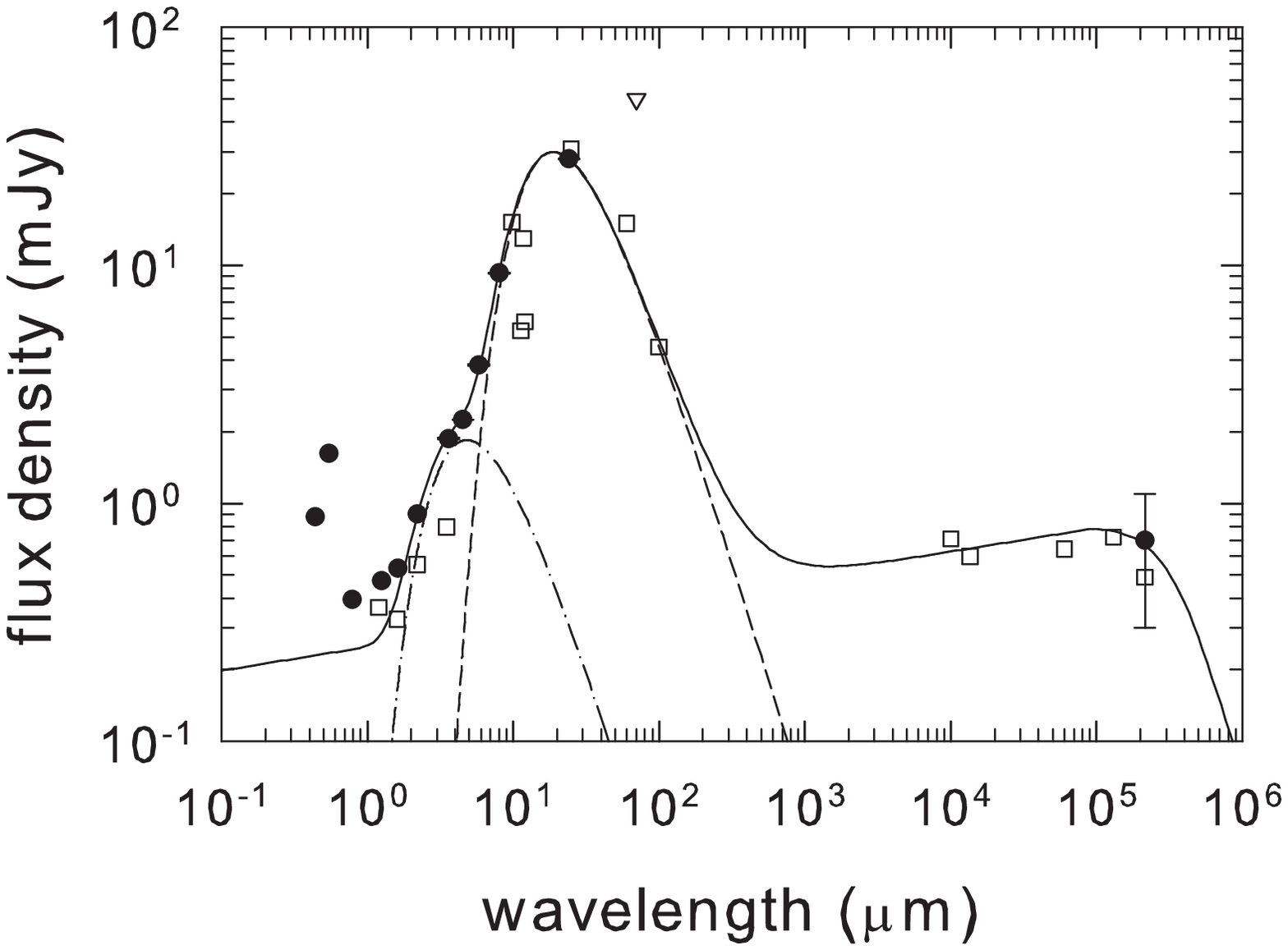}}
\figurecaption{3.}{SED of SMP~24 from B band to radio frequencies. Dashed line represents the best fit to the FIR band with the black body (BB) approximation. The dot-dashed line represents the best fit (BB approximation) to the empirical distribution in NIR and MIR bands. The summed SED of the radio-continuum, dust and the hot dust is plotted with solid line.  The triangle represents the detection limit in the 70~$\mu$m band. Overplotted boxes represent the observed SED of a young Galactic PN IC~418, scaled to the distance of the SMC (see text for more details).}

\vfill\eject

{\ }

% Serbian abstract

% Title

\naslov{\textrm{SMC~SMP~24}: NOVA RADIO PLANETARNA MAGLINA U MALOM MAGELANOVOM OBLAKU}

% Authors

\authors{I.~S.~Boji\v{c}i\'c, M.~D.~Filipovi\'c and  E. ~J.~Crawford}

\vskip3mm

% Address

\address{University of Western Sydney, Locked Bag 1797, Penrith South DC, NSW 1797, Australia}
\Email{i.bojicic}{uws.edu.au}

\vskip.7cm

% UDC

\centerline{UDK \udc}

% Papertype

\centerline{\rit Originalni nauqni rad}

\vskip.7cm

\begin{multicols}{2}
{

% Abstract

\rrm 
U ovoj studiji predstav{lj}amo nove {\rm ATCA} rezultate posmatra{nj}a u radio-kontinumu vangalaktiqke planetarne magline: \textrm{SMC~SMP~24}. Nova radio posmatranja su konzistentna sa posmatra{nj}ima na ostalim talasnim du{\zz}inama i parametrima na{dj}enim u dosadax{nj}im istra{\zz}va{nj}ima. \textrm{SMC~SMP~24} je najverovatnije mlada i kompaktna planetarna maglina, optiqki neprobojna na frekvencijama ispod \textrm{2~GHz.}}
\end{multicols}

\end{document}